\documentclass[prl,a4paper,twocolumn,superscriptaddress,amsfonts,amssymb,amsmath,floatfix]
{revtex4}
\usepackage{graphicx}
\usepackage{subfigure}
\usepackage{times}
\usepackage{mathrsfs}

\begin{document}

\author{Sananda Biswas} 
 \affiliation{The Abdus Salam International Centre for Theoretical Physics, Strada Costiera 11, Trieste, Italy} 
\affiliation{Theoretical Sciences Unit, Jawaharlal Nehru Centre for Advanced Scientific Research, Jakkur, Bangalore, India}
\author{Shobhana Narasimhan} 
\affiliation{Theoretical Sciences Unit, Jawaharlal Nehru Centre for Advanced Scientific Research, Jakkur, Bangalore, India}
\date{\today}

\begin{abstract}
Predicting the ground state of alloy systems is challenging due to the large number of possible 
configurations. We identify an easily computed descriptor for the stability of binary surface 
alloys, the effective coordination number $\mathscr{E}$. We show that $\mathscr{E}(M)$ 
correlates well with the enthalpy of mixing, from density functional theory (DFT) calculations on  
$M_x$Au$_{1-x}$/Ru [$M$  = Mn or Fe]. At each $x$, the most favored structure has the 
highest [lowest] value of $\mathscr{E}(M)$ if the system is non-magnetic [ferromagnetic]. 
Importantly, little accuracy is lost upon replacing $\mathscr{E}(M)$ by $\mathscr{E}^*(M)$, 
which can be quickly computed without performing a DFT calculation, possibly offering a simple 
alternative to the frequently used cluster expansion method.

\end{abstract}

\title{A simple descriptor and predictor for the stable structures of two-dimensional surface alloys}
\maketitle

\section{Introduction}

The goal of rational materials design is the {\it in silico} engineering of novel materials which 
possess optimal properties for specific applications \cite{hafner2006}. However, this still  
remains a daunting task; a significant obstacle is that we do not, at present, have any simple way  
of predicting the structure that corresponds to the global minimum in the configurational space 
of a material with a given chemical composition. One therefore has to examine a large number of  
candidate structures before one can be reasonably confident that one has found the structure 
corresponding to the ground state. This procedure is time-consuming, both because of the very 
large number of configurations that have to be screened, and because each individual calculation 
(typically performed using {\it ab initio} density functional theory) can itself be computationally 
demanding.

In order to save computational time, one therefore looks for `descriptors'
\cite{curtaroloNM2013}. A good descriptor is a quantity that is easily computed, yet correlates 
well with the property of interest (e.g., the structural stability). Moreover, to be truly useful, a 
descriptor should ideally also function as a {\it predictor}, i.e., one should be able to foretell the 
properties of a new system with high accuracy, without having to perform the {\it {ab initio}} 
calculation at all. Though certain descriptors also incorporate a degree of physical insight (e.g., 
the $d$-band center \cite{hammerNature1995} or the generalized coordination number
\cite{sautetACIE2014}, both of which have been shown to correlate well with catalytic activity), 
in more complex cases, it is difficult to attribute physical interpretations to the form of the 
descriptor. For instance, it is possible to systematically build up multi-dimensional descriptors for 
various properties of materials using machine learning algorithms. This has been shown, e.g., by 
Ghiringhelli {\it et al.} for the crystal structures of binary octet compounds \cite{ghiringhelliPRL2015}.

One area of condensed matter physics which would benefit greatly from efforts to develop such 
descriptors is the study of alloys. This could offer an alternative to the cluster expansion 
method (CEM), which has hitherto been perhaps the most promising approach toward tackling 
the challenging problem of computing stable alloy phases \cite{sanchez1,sanchez2,lakPRb1992}. 
In the CEM, a set of dominant interactions is determined by fitting to a database of 
first-principles results; this expansion is then used to extrapolate to much larger unit cells. The 
advantage of the CEM is that it can save considerable time (compared to a full-fledged DFT 
calculation), the disadvantage is that there is no general rule about how many or which terms in 
the expansion should be retained, and large databases may be necessary. There have also been 
several attempts by previous authors to rationalize the phase diagrams of bulk alloys in terms of 
a few simple parameters \cite{phillipsRMP1970, vanVechtenPR1968, zungerPRB1980, 
pettiforSSC1984, saadPRB2012}, with mixed success. Here we consider the somewhat more 
tractable problem of two-dimensional binary surface alloys. We show that a single easily 
computed descriptor, the `effective coordination number' $\mathscr{E}$, correlates well with 
the energetics of various alloy configurations. Interestingly, we show that the most favored value 
of $\mathscr{E}$ flips, depending on whether  the system is non-magnetic or ferromagnetic. 
Moreover, we show that though $\mathscr{E}$ requires knowledge of the relaxed structure 
(which, in principle, would be known only at the end of a DFT calculation), it can be replaced, with 
little loss of predictive power, by another quantity $\mathscr{E}^{*}$ which makes use of 
unrelaxed geometries.

As proof-of-concept, we present results on four systems: (i) non-magnetic and (ii) 
ferromagnetic states of two binary systems: (a) Fe-Au and (b) Mn-Au, always on a Ru(0001) 
substrate.

Surface alloys are systems with two or more elements forming a two-dimensional alloy on a 
substrate \cite{nielsenPRL1993}. These alloys display atomic level mixing in the surface layer; 
magnetic surface alloys are particularly appealing, both because it has been shown theoretically
that exchange interactions can contribute significantly to miscibility 
\cite{mehendalePRL2010, marathePRB2013}, and because of the possibility of using them in a 
variety of applications related to magnetic memory storage and spintronics. The reduced 
symmetry in such a system could conceivably lead to a variety of interesting properties, such as 
enhanced magnetic moments and/or an increase in the magnetic anisotropy energy. It is 
especially intriguing to note that it is possible to form surface alloys out of elements that are 
immiscible in the bulk \cite{nielsenPRL1993, stevensPRL1995}. Of the binary systems 
considered in this study, we remark that Fe and Au are essentially immiscible in the bulk, though 
they have been shown to form a stable surface alloy with long range order when co-deposited 
on Ru(0001); however, Mn and Au do form stable bulk alloy 
phases~\cite{massalskiBAPD1985, udvardiPRB2006}.

Our calculations have been performed using either spin-polarized or non-spin-polarized density 
functional theory (DFT) as implemented in the Quantum ESPRESSO package \cite{espresso},  
which uses a plane-wave basis set. The plane-wave cutoffs for the expansion of the electronic 
wavefunctions and the related charge densities were set as 40 and 400 Ry, respectively. 
Ultrasoft pseudopotentials \cite{uspp} were used to describe the ion-electron interactions, while    
the exchange-correlation functional was treated within a generalized gradient approximation of 
the Perdew-Burke-Ernzerhof form \cite{PBE}.

In this paper, we restrict ourselves to considering pseudomorphic surface alloys, by which we 
mean that the atomic density in the surface layer is equal to that in the substrate layers.  This 
approximation is expected to be valid for both systems considered by us, since the bulk nearest 
neighbor (NN) distance of the Ru substrate lies in between that of Au and that of either Fe or Mn.  
However, we can either choose to permit in-plane as well as out-of-plane relaxations of the atoms 
in the surface layer (we will refer to this as PR, for pseudomorphic relaxed), or force the surface 
atoms to remain at the hexagonal close packed (hcp) sites on the Ru(0001) surface (PU, for 
pseudomorphic unrelaxed). Note that the PR positions can only be obtained by performing an 
{\it ab initio} DFT calculation where the geometries are relaxed making use of Hellmann-Feynman 
forces, whereas  no DFT calculation is necessary to obtain a PU structure. In the PR calculations, 
the threshold for convergence of forces was kept at 0.025 eV/\AA.

For every surface alloy composition $M_x$Au$_{1-x}$ (where $M$ = Fe or Mn) an infinite 
number of atomic configurations is possible, with varying numbers of atoms in the surface unit 
cell. As already mentioned above, this is what makes the problem of finding the most stable 
structure by a search through all possible configurations intractable. We therefore first generate 
all possible structures containing one to five atoms per surface unit cell, and several containing 
six atoms; in this way, we obtain 41 distinct symmetry-inequivalent periodic structures 
\cite{marathePRB2013}. For each structure, we both `permit' magnetism (by performing a spin 
polarized DFT calculation) and `suppress' magnetism (by performing a non-spin-polarized 
calculation). We divide these 41 structures, according to their structural similarities, into  nine 
groups (A)--(H) [see supplementary material]. We note that both in-plane and out-of-plane 
relaxations differ significantly in the non-magnetic (NM) and ferromagnetic (FM) cases, with a 
tendency for neighboring $M$ atoms to cluster together being observed for NM configurations.

\begin{figure}[!t]
\centering
\setlength\fboxsep{0pt} 
\setlength\fboxrule{0pt}
\includegraphics[width=7.7cm]{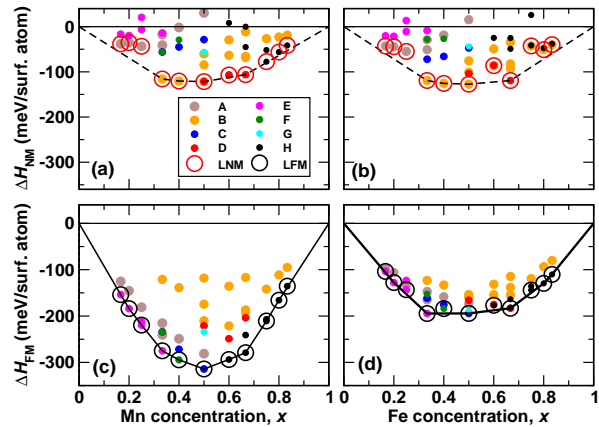}
\caption{(Color online) Enthalpy of mixing per surface atom, $\Delta H$,  of $M_x$Au$_{1-x}$/Ru(0001) as a 
function of $x$ the concentration of $M$ atoms, where $M$= Mn or Fe. (a) and (b) show  
results for the non-magnetic (NM) phases, (c) and (d) for the ferromagnetic (FM) phases.  The 
configurations have been grouped according to their structural similarities, with each group 
denoted by a different color [see supplementary material for the structures in groups (A)-(H)]. The black dashed and solid lines indicate the NM and FM 
convex-hulls, respectively. The points corresponding to the lowest energy NM and FM structures 
at each $x$ are circled in red and black, respectively. }
\label{fig:deltaH}
\end{figure}

For all the structures considered here, we find that the FM state is lower in energy than the NM 
one. For both the FM and NM states, the enthalpy of mixing $\Delta H$ for the surface alloy, 
relative to the phase segregated components $M$/Ru and Au/Ru, is given by:
\begin{eqnarray}\label{eq:delta_H}
\Delta H &= E({M}_x{\rm Au}_{1-x}/{\rm Ru}) - x E({M/\rm Ru})\nonumber \\
& -(1-x)  E({\rm Au/Ru}),
\end{eqnarray}
where $E(X)$ is the total energy of system $X$, and the first two terms on the right-hand-side 
are both evaluated in the corresponding magnetic state, while the third term is always 
non-magnetic. For a given surface alloy to be stable, a necessary but not sufficient condition is 
that $\Delta H < 0$.

In Figs.~\ref{fig:deltaH}(a) -- (d), we have plotted the enthalpy of mixing 
$\Delta H_{\rm{NM}}$ and $\Delta H_{\rm{FM}}$, for all the structures in the NM and the FM 
states, respectively [we note that the results in Figs.~\ref{fig:deltaH}(b) \& (d) 
are very similar to those previously published in Ref.~\onlinecite{mehendalePRL2010}]. These 
graphs have several noteworthy features: (i) though $\Delta H_{\rm{FM}} < 0$ always, in some 
cases $\Delta H_{\rm{NM}} > 0$, i.e., in some cases the stability is provided solely by exchange 
interactions, (ii) $|\Delta H_{\rm{FM}}| \gg |\Delta H_{\rm{NM}}| $, i.e., even in the other 
cases, a large part of the stability arises from magnetism 
\cite{mehendalePRL2010, marathePRB2013}, and (iii) of particular interest for the issues we 
wish to focus on in this paper, in general, the lowest energy configurations (see the colors of the 
circled dots) differ in the NM and FM cases (except for a few configurations at large $x$).

To determine which of these lowest energy configurations of $M_x$Au$_{1-x}$/Ru(0001) are 
stable with respect to phase segregation, we have computed the NM and FM convex hulls [shown 
by the dashed and solid lines, respectively, in Figs.~\ref{fig:deltaH}(a) -- 
(d)]. We note that, in general, the stable configurations are different for the NM and FM cases, 
for both Mn-Au/Ru and Fe-Au/Ru. Our result from DFT that the most stable alloy phase for the 
Fe-Au system is a $\sqrt{3}\times\sqrt{3}$ structure at $x=0.33$ is in agreement with 
experiment \cite{mehendalePRL2010}. For Mn-Au, our finding that the most stable phase occurs 
at $x=0.5$ and has a $2\times\sqrt{3}$ unit cell is a theoretical prediction awaiting 
experimental confirmation.

\begin{figure}[!t] 
\setlength\fboxrule{0pt}
\centering
\includegraphics[width=7.7cm]{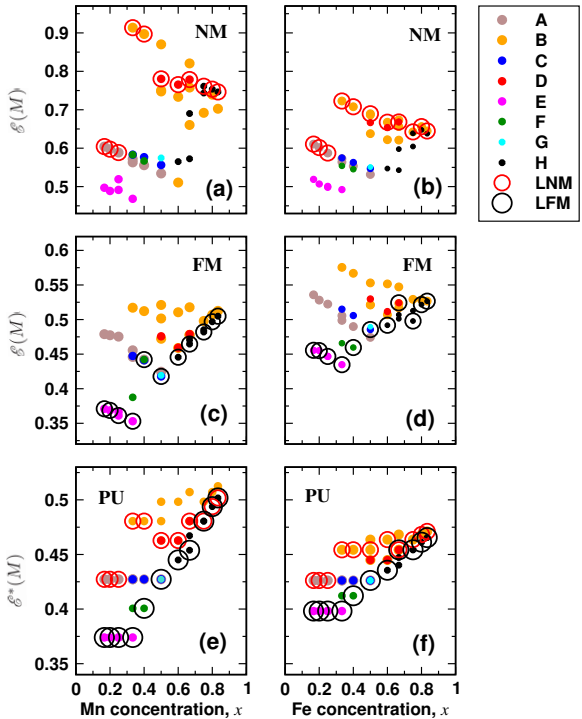}
\caption{(Color online) Value of the descriptor $\mathscr{E}$ for all 41 configurations considered, as a function of $x$, the concentration of Mn or Fe atoms, for (a) -- (c) Mn$_x$Au$_{1-x}$/Ru(0001) and (d) -- (f) Fe$_x$Au$_{1-x}$/Ru(0001). See supplementary material for the structures in groups (A)-(H). NM, FM and PU refer to the non-magnetic, ferromagnetic and pseudomorphic unrelaxed cases, respectively. Configurations have been grouped according to their symmetry, with each group indicated by a different color. The points corresponding to the lowest energy NM and FM configurations have been circled in red and black, respectively. Note that in Fig.~(e) the red and black circles coincide for $x \geq 0.75$, whereas in Fig.~(f), they coincide only for $x = 0.67$. }
\label{fig:ecn}
\end{figure}

It is known that the stability of a surface alloy is determined mainly by two factors: the band 
energy and (if magnetic) the exchange energy terms \cite{bluegelAPA1996}; these in turn are 
sensitive to how many neighbors of each type every surface atom has, and how far away they 
are. In order to characterize these, simply counting the number of nearest neighbor atoms 
would not suffice, yielding a nominal coordination number of nine for any atom on a hcp(0001) 
surface. Instead, we define the effective coordination number of an atom $i$ as
\cite{baraldiNJP2007}:
\begin{equation}\label{eq:ecn}
\mathscr{E}(i)  = \frac{\displaystyle \sum_{j}^{} \rho{_{T_j}}(r_{ij})}{\displaystyle \sum_{j}^{} \rho_{T_{i=j}}^{\rm{bulk}}(r_{ij})},
\end{equation}
where in the numerator, the sum is taken over all the neighboring atoms $j$, of type $T_j$, at 
a distance $r_{ij}$ from the $i$th atom in the surface layer. The denominator is evaluated in 
the bulk structure of the atom $i$, the sum is taken over all the neighboring atoms $j$, of type 
$T_{j=i}$. $\rho{_{T_j}}(r)$ is the atomic charge density at a distance $r$ from the nucleus of 
an isolated atom of type $T_j$. When computing the denominator in Eq.~\ref{eq:ecn}, we have 
considered the $\alpha$-Mn structure for Mn, the bcc structure for Fe, and the fcc structure for 
Au. In those two-dimensional surface alloy structures that contain more than one kind of 
symmetry-inequivalent atom of a species $i$, we compute $\mathscr{E}(i)$ by taking an 
appropriately weighted average over the different kinds of $i$ atoms. $\mathscr{E}(i)$ reflects 
the ambient electron density at the site of the atom $i$, in the spirit of the embedded atom 
model or effective medium theory \cite{dawPRL1983, dawPRB1989, jacobsenPRB1987}.

In Figs.~\ref{fig:ecn} (a) -- (d) , we plot the values of $\mathscr{E}(M)$ for 
the PR structures of all the 41 configurations considered by us; note the clustering of points 
that belong to the same group of structures. Importantly, we see that for the two NM systems, 
the lowest energy structure is almost always the structure with the {\it{largest}} value of 
$\mathscr{E}(M)$ [see the positioning of the red circles at each value of $x$].  However, the 
situation is exactly the opposite when we go to the FM cases: the lowest energy structure is 
almost always the one with the {\it{smallest}} value of $\mathscr{E}(M)$ [see the black 
circles]. Moreover, we find that at each value of $x$, in general, as $\mathscr{E}(M)$ increases, 
$\Delta H_{NM}$ decreases, and $\Delta H_{FM}$ increases. On the whole, $\mathscr{E}(M)$ 
acts as an excellent descriptor, indicating not just the most stable structure of the surface alloy, 
but also the relative ordering of the enthalpy of formation of different configurations, for both 
the magnetic and non-magnetic cases.

Not only is $\mathscr{E}(M)$ simple to compute, it also lends itself easily to physical
interpretation, and provides an ideal tool for examining two of the principal interactions that 
compete in determining the stability of surface alloys, viz., the band energy $E_b$ and magnetic 
energy $E_m$ \cite{bluegelAPA1996}. When $\mathscr{E}(M)$ is large, hybridization between 
$M$ atoms is enhanced, and the bandwidth is increased. As a result, 
$E_b \equiv \int{_{-\infty}^{E{_{F}}=0}} \epsilon n(\epsilon)\ d\epsilon$ becomes large, where 
$\epsilon$ runs over the electronic energies, $n(\epsilon)$ is the electronic density of states, 
and $E_F$ is the Fermi energy (set equal to 0). Note that with this definition, $E_b < 0$. Note 
also that a large $\mathscr{E}(M)$ would imply a broad and low $n(\epsilon)$, and thus a low 
value of  $n_{\rm{NM}}(E_F)$, the non-magnetic density of states at the Fermi energy. 
According to the Stoner model \cite{stonerPRS}, this would disfavor ferromagnetism. In contrast, 
when $\mathscr{E}(M)$ is low, the bandwidth is decreased, and exactly the reverse arguments 
apply. Thus, in such a situation, $E_b$ would be small and $E_m=-Im^2$ would tend to be large, 
where $I$ is the Stoner parameter, and $m$ is the magnetic moment.  Thus, our findings should 
hold for systems where the stability of surface alloys is dominated by either band energy or
exchange energy terms. It remains to be verified whether this continues to hold true, e.g., when 
magnetic moments on the $M$ atoms are small, or elastic effects are strong.

An important question remains: can one predict the lowest energy configuration without doing 
the DFT calculations? Note that while calculating $\mathscr{E}(M)$ to obtain the points in
Figs.~\ref{fig:ecn} (a) -- (d), we made use of fully relaxed atomic PR coordinates, which 
one can obtain only after performing DFT calculations. We now check what would happen if we 
instead used the PU coordinates. The effective coordination number thus obtained is defined as $
\mathscr{E}^{*}$ and is plotted in Figs.~\ref{fig:ecn} (e) and (f).  We see that for the 
most part, the correct lowest energy structure is predicted even using $\mathscr{E}^{*}(M)$.

\begin{figure}[t]
\includegraphics[width=8.3cm]{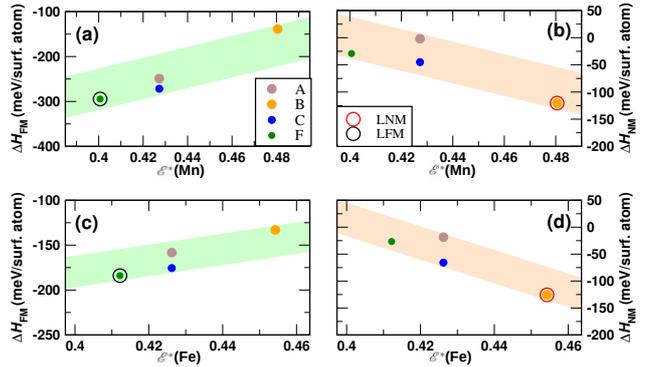}
\caption{(Color online) $\mathscr{E}^{*} $ works as a predictor. There is a correlation between $\Delta H$ (in meV per surface atom) and the descriptor $\mathscr{E}^{*} $, obtained simply from PU configurations without DFT calculations. As an example, results are shown at $x=0.4$ for (a) FM Mn-Au, (b) NM Mn-Au, (c) FM Fe-Au, and (d) NM Fe-Au, all on Ru(0001). The shaded areas are a guide to the eye, indicating both trends and spreads in values. The lowest energy configurations in the FM and NM cases are labelled as LFM (black open circles) and LNM (red open circles), respectively.}    
\label{fig:deltaH_vs_ecn}
\end{figure}

As a test of the performance of $\mathscr{E}^{*}(M)$ as a predictor, in 
Fig.~\ref{fig:deltaH_vs_ecn} we have plotted four graphs; as the abscissa we have 
$\mathscr{E}^{*}(M)$, and as the ordinate we have $\Delta H$ as calculated from DFT. If 
$\mathscr{E}^{*}(M)$ is a good predictor, it should correctly identify the configuration with the 
lowest $\Delta H$. We see that for the sample case that we have plotted 
$(x = 0.4)$, $\mathscr{E}^{*}(M)$ correctly predicts the lowest energy structure in all four
cases, and further, there is a reasonably good correlation between the values of ordinate and 
abscissa, with only small deviations from monotonicity. There are two ways in which 
$\mathscr{E}^{*}$ can `underperform': the lowest energy structure may not correspond to an 
extremal value of $\mathscr{E}^{*}(M)$, though it may lie close in energy; alternatively, more 
than one structure could conceivably correspond to the extremal value of $\mathscr{E}^{*}(M)$, 
because of the high symmetry of the PU structures. However, one can still save considerable 
computational time by first generating PU configurations, calculating $\mathscr{E}^{*}(M)$, thus 
identifying a pool of `best' candidate structures (the ones that are predicted to be the lowest 
few in energy), and then performing {\it {ab initio}} calculations on only this reduced pool of 
structures.

One can also try to understand why, in some cases, $\mathscr{E}^{*}(M)$ predicts the ground 
state structure incorrectly. For example in Fig.~\ref{fig:ecn}(e) and (f), at $x=0.5$, we 
find that the NM configuration with the highest $\mathscr{E}^{*}(M)$ has a rather high and 
anisotropic stress, and the system instead chooses to form in a structure which, while having a 
reasonably high $\mathscr{E}^{*}(M)$, has an isotropic surface stress.

In summary, we suggest that a quantity that is easy to compute, the effective coordination 
number $\mathscr{E}$, serves as a good descriptor for the stability of binary surface alloys on 
a substrate. For magnetic surface alloys, the lowest energy configuration is almost always the 
one with the lowest value of $\mathscr{E}(M)$; in contrast, in the non-magnetic case, the most 
favored configuration tends to have the highest value of $\mathscr{E}(M)$. We note that
though in the systems studied in this paper, $M$ is always the `magnetic' element, for a general 
binary alloy comprised of non-magnetic atoms, $M$ will correspond to the atom with the smaller 
atomic radius.

Further, even if one instead calculates $\mathscr{E}^{*}(M) $ using `pseudomorphic unrelaxed' 
coordinates (sidestepping the need for an {\it {ab initio}} calculation) one can predict, with a 
high success rate, the configuration which will have the lowest energy. The success rate can be 
further improved by extending the pool of candidate structures to include the few configurations 
which have the highest (for NM) or lowest (for FM) values of $\mathscr{E}^{*}(M)$ and then 
performing first-principles calculations only on this considerably reduced number of structures. 
Here, we have demonstrated that the above statements hold for four model systems (two 
magnetic and two non-magnetic). The ease of computation of ${\mathscr{E}^*}$ allows one to 
greatly expand the space of alloy configurations searched, at minimal additional cost, thus
making it ideal for incorporation within high-throughput programs of computational materials 
discovery targeted at specific applications.

\section*{Acknowledgement}
\noindent
We thank M.~Marathe, S.~Rousset and  V.~Repain for many helpful discussions, and acknowledge funding from CEFIPRA, the Indo-French Centre for the Promotion of Advanced Research.

\include{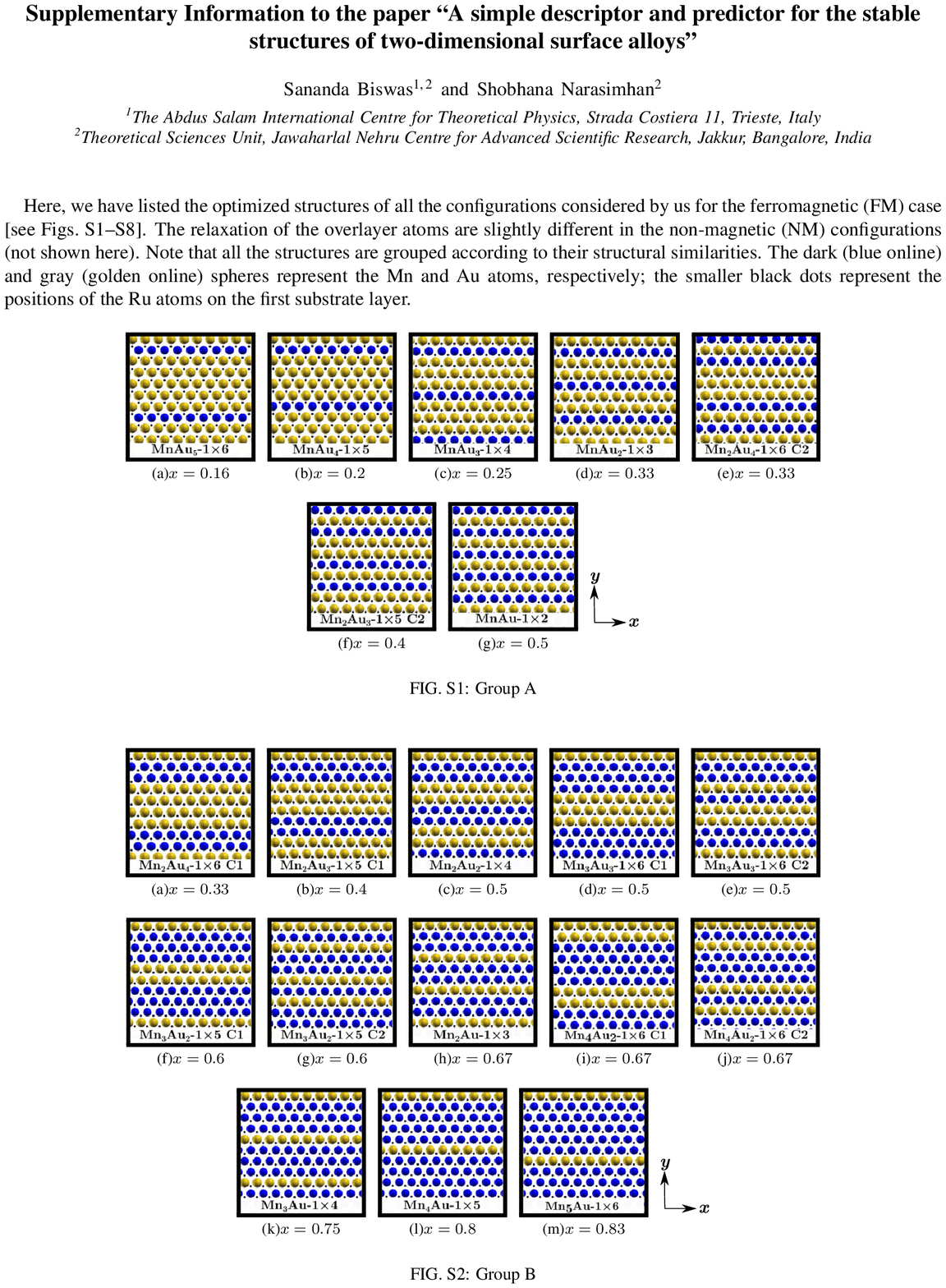}
\end{document}